\documentclass{PoS}

\usepackage{amssymb,graphicx,amsmath,array,verbatim,setspace}

\newcommand{\beq}{\begin{eqnarray}}
\newcommand{\eeq}{\end{eqnarray}}
\newcommand{\p}{\partial}

\newcommand{\bpm}{\begin{pmatrix}}
\newcommand{\epm}{\end{pmatrix}}
\newcommand{\Z}{\mathbb{Z}}

\newcommand{\C}{\mathbb{C}}

\newcommand{\ba}{\left(\begin{array}}
\newcommand{\ea}{\end{array} \right)}

\title{Lattice study on the twisted ${\mathbb C}P^{N-1}$ models on ${\mathbb R}\times S^{1}$\,\thanks{This work is supported by MEXT-Supported Program ``Topological Science" S1511006
and by JSPS KAKENHI 18H01217, 19K03875, 18K03627, 19K03817, 16H03984, 15H05855. 
Numerical simulations were performed on SX-ACE, Osaka University and TSC, Keio University.}}

\ShortTitle{Lattice study on the twisted ${\mathbb C}P^{N-1}$ models on ${\mathbb R}\times S^{1}$}

\author{\speaker{Tatsuhiro Misumi}\\
        Department of Mathematical Science, Akita University\\
Research and Education Center for Natural Sciences, Keio University\\
iTHEMS, RIKEN\\
        E-mail: \email{misumi@phys.akita-u.ac.jp}}

\author{Toshiaki Fujimori\\
        Department of Physics, Research and 
Education Center for Natural Sciences, 
Keio University\\
        E-mail: \email{toshiaki.fujimori018@gmail.com}}

\author{Etsuko Itou\\
        Research and Education Center for Natural Sciences, 
Keio University\\
Department of Mathematics and Physics, Kochi University\\
Research Center for Nuclear Physics (RCNP), Osaka University,\\
        E-mail: \email{ itou@yukawa.kyoto-u.ac.jp}}

\author{Muneto Nitta\\
        Department of Physics, Research and 
Education Center for Natural Sciences, 
Keio University\\
        E-mail: \email{nitta@phys-h.keio.ac.jp}}

\author{Norisuke Sakai\\
        Department of Physics, Research and 
Education Center for Natural Sciences, 
Keio University\\
        E-mail: \email{norisuke.sakai@gmail.com }}

\abstract{We report the results of the lattice simulation of the $\C P^{N-1}$ sigma model 
on $S_{s}^{1}$(large) $\times$ $S_{\tau}^{1}$(small).
We take a sufficiently large ratio of the circumferences to approximate the model on $\mathbb R \times S^1$. 
For periodic boundary condition imposed in the $S_{\tau}^{1}$ direction, 
we show that the expectation value of the Polyakov loop undergoes a deconfinement crossover as the compactified circumference is decreased, where the peak of the associated susceptibility gets sharper for larger $N$.
For ${\mathbb Z}_{N}$ twisted boundary condition, we find that, even at relatively high $\beta$ (small circumference), the regular $N$-sided polygon-shaped distributions of Polyakov loop leads to small expectation values of Polyakov loop, 
which implies unbroken ${\mathbb Z}_{N}$ symmetry if sufficient statistics and large volumes are adopted.
We also argue the existence of fractional instantons and bions by investigating the dependence of the Polyakov loop on $S_{s}^{1}$ direction, which causes transition between ${\mathbb Z}_{N}$ vacua.
}

\FullConference{37th International Symposium on Lattice Field Theory - Lattice2019\\
		16-22 June 2019\\
		Wuhan, China}

\begin{document}

\section{Introduction}
Although non-perturbative properties of the ${\mathbb C}P^{N-1}$ model on $\mathbb{R}^2$ have been long studied by analytical and numerical techniques including lattice simulations \cite{Campostrini:1992ar, Farchioni:1993jd, Fujimori:2019skd},
the model on $\mathbb{R} \times S^1$ (or at finite temperature) has not been well investigated.
The recent large-$N$ analyses of the model on $\mathbb{R} \times S^1$ 
with periodic boundary conditions (PBC) have been attracting much attention \cite{Monin:2015xwa,Bolognesi:2019rwq}.
One of the questions we focus on here is how the expectation value of the order parameter for confinement-deconfinement depends on the compactification circumference for PBC.
On the other hand, the model on $\mathbb{R} \times S^1$ with $\Z_{N}$ twisted boundary conditions (${\mathbb Z}_{N}$-TBC) also attracts a lot of attention, where the $\Z_{N}$ symmetry is exact.
The questions whether it undergoes a phase transition or has adiabatic continuity \cite{Dunne:2012ae,Sulejmanpasic:2016llc} and whether fractional instantons have consequences \cite{Eto:2006mz} are of great importance in terms of the resurgence theory \cite{Dunne:2012ae,Misumi:2014jua,Fujimori:2016ljw}.

In this report, we investigate the ${\mathbb C}P^{N-1}$ model on $S_{s}^{1}$(large) $\times$ $S_{\tau}^{1}$(small) with a sufficiently large ratio of the circumferences by lattice Monte Carlo simulations to answer the above questions. 
We mainly focus on the distribution and expectation values of Polyakov loop. 
Our results are summarized as follows.
(1) For PBC, by adopting the absolute value of the expectation value of the Polyakov loop as a confinement-deconfinement order parameter, we find that its dependence on the compactification circumference $L_{\tau}$ exhibits a crossover behavior and the peak of its susceptibility gets sharper with $N$ increases \cite{Fujimori:2019skd}. 
(2) For ${\mathbb Z}_{N}$-TBC, we find that, even at relatively high $\beta$ (small $L_{\tau}=1/T$), the regular $N$-sided polygon-shaped distributions of Polyakov loop give the small expectation values, which imply unbroken ${\mathbb Z}_{N}$ symmetry if sufficient statistics and large volumes are adopted.
We also argue the existence of fractional instantons and bions.

\section{Lattice setup}\label{sec:setup}
Let us begin with the continuum action of the ${\mathbb C}P^{N-1} = 
{\rm SU}(N)/({\rm SU}(N-1)\times {\rm U}(1))$ sigma models,
which is given as $S={1\over{g^{2}}} \int d^2 x D_\mu \bar\phi D_\mu \phi$ in terms of 
an $N$-component complex scalar field $\phi$ 
with the constraint $|\phi|^2 =1$ and an auxiliary $U(1)$ gauge 
field $A_\mu$ with the covariant derivative $D_\mu \phi= (\p_\mu + iA_\mu) \phi$. 
The corresponding lattice action \cite{Campostrini:1992ar, Farchioni:1993jd, Fujimori:2019skd}
is
\begin{equation}
S=N\beta \sum _{n,\mu} \left( 2- \bar{\phi}_{n+\hat{\mu}}\cdot \phi_{n} \,
\lambda_{n,\mu} - \bar{\phi}_{n}\cdot \phi_{n+\hat{\mu}} 
\bar{\lambda}_{n,\mu} \right)\,,
\label{eq:latt-action}
\end{equation}
with $\bar{\phi}_{n} \cdot \phi_{n} =1$ and $\lambda_{n,\mu}$ 
being a link variable corresponding to the auxiliary U($1$) gauge field. 
$n=(n_x, n_\tau)$ labeling sites on the lattice 
run as $n_x = 1, \cdots, N_s$ and $n_\tau = 1, \cdots, N_\tau$
and $N\beta$ corresponds to the inverse of the bare coupling $1/ g^{2}$.
$S_s^{1}$ and $S^1_\tau$ have circumferences $L_{s} = N_{s} a$ and $L_{\tau} = N_{\tau} a$, respectively.
As discussed in \cite{Fujimori:2019skd}, the renormalization group gives the relation 
between the lattice parameter $\beta$ and the lattice spacing $a$ via the lattice scale $\Lambda_{lat}$ as
$a\Lambda_{lat}  = {1 \over{\sqrt{32}}} (2\pi\beta)^{2\over{N}} 
e^{-2\pi\beta -{\pi\over{2N}}}$.
For our purpose we adopt the condition $L_s \gg L_\tau$, which enables us to approximately simulate the model on $\mathbb R \times S^1$. 
Here $L_{\tau}$ corresponds to an inverse temperature $1/T$ and smaller $L_{\tau}$ or higher $\beta$ with fixed $N_\tau$ corresponds to higher $T$.

\section{Polyakov loop for periodic boundary conditions}\label{sec:Ploop}

In this section, we show the results for PBC, with emphasis on the the expectation value of Polyakov loop as an order parameter of confinement-deconfinement.
The arguments of this section are based on the work \cite{Fujimori:2019skd} by the present authors.
The lattice size in this section is mainly $(N_{s},N_{\tau})=(200,8)$ but we vary $N_{s}$ between $40$ and $200$ to study the finite-volume effects. We work on the values of parameters as $N=3,5,10,20$ and $0.1 \leq \beta \leq 3.9$.

On ${\mathbb R}^{2}$ the expectation value of Wilson loop $\langle W(C) \rangle$, which depends on the string tension $\sigma$ as $\approx e^{-\sigma RT}$ with $R$, $T$ being space and Euclidean-time lengths of $C$,
characterizes the confinement of electrically charged particles \cite{Campostrini:1992ar}.
For compactified theory with PBC, the Wilson loop is expressed as a correlator of Polyakov loops $P(x) \equiv {\mathcal P} \exp(i\int_0^{L_\tau} d\tau A_\tau)_{x}$ as $\langle P(R) P^\dag (0) \rangle$. 
With taking account of the clustering property in $R \to \infty$ limit, one finds that the expectation value of Polyakov loop should vanish $\langle P \rangle=0$ in the confinement phase with $\sigma\not=0$. 
Thus, $\langle P \rangle$ can be adopted as the order parameter of confinement-deconfinement in the $L_\tau \ll L_s$ system.
It is also an approximate order parameter of ${\mathbb Z}_{N}$ symmetry although the symmetry is not exact for the model on ${\mathbb R}\times S^{1}$ with PBC.
We note that it is an exact order parameter for ${\mathbb Z}_{N}$-TBC since the symmetry is exact.

\begin{figure}[t]
\centering
 \includegraphics[width=0.7\linewidth]{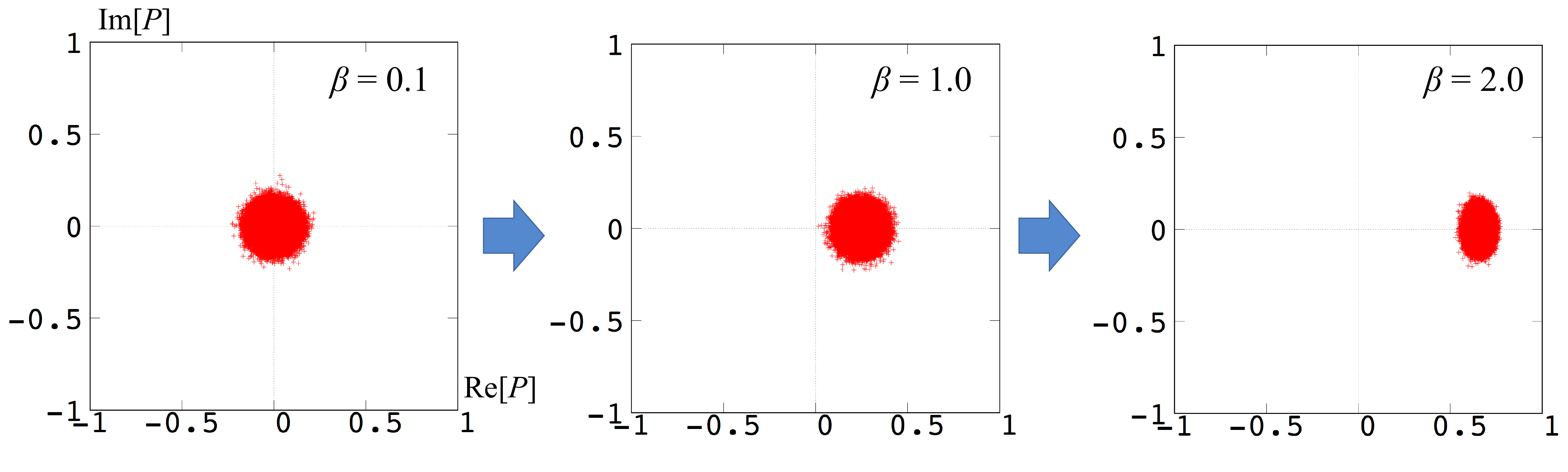}
 \caption{Distribution plots of Polyakov loop for $\beta=0.1$(left), $\beta=1.0$(center) and $\beta=2.0$(right) for $N=3$ with PBC. The Polyakov loops are distributed around the origin at low $\beta$ and gradually leave from the origin at higher $\beta$.}
\label{fig:dist}
\end{figure}

\begin{figure}[t]
\centering
 \includegraphics[width=0.55\linewidth]{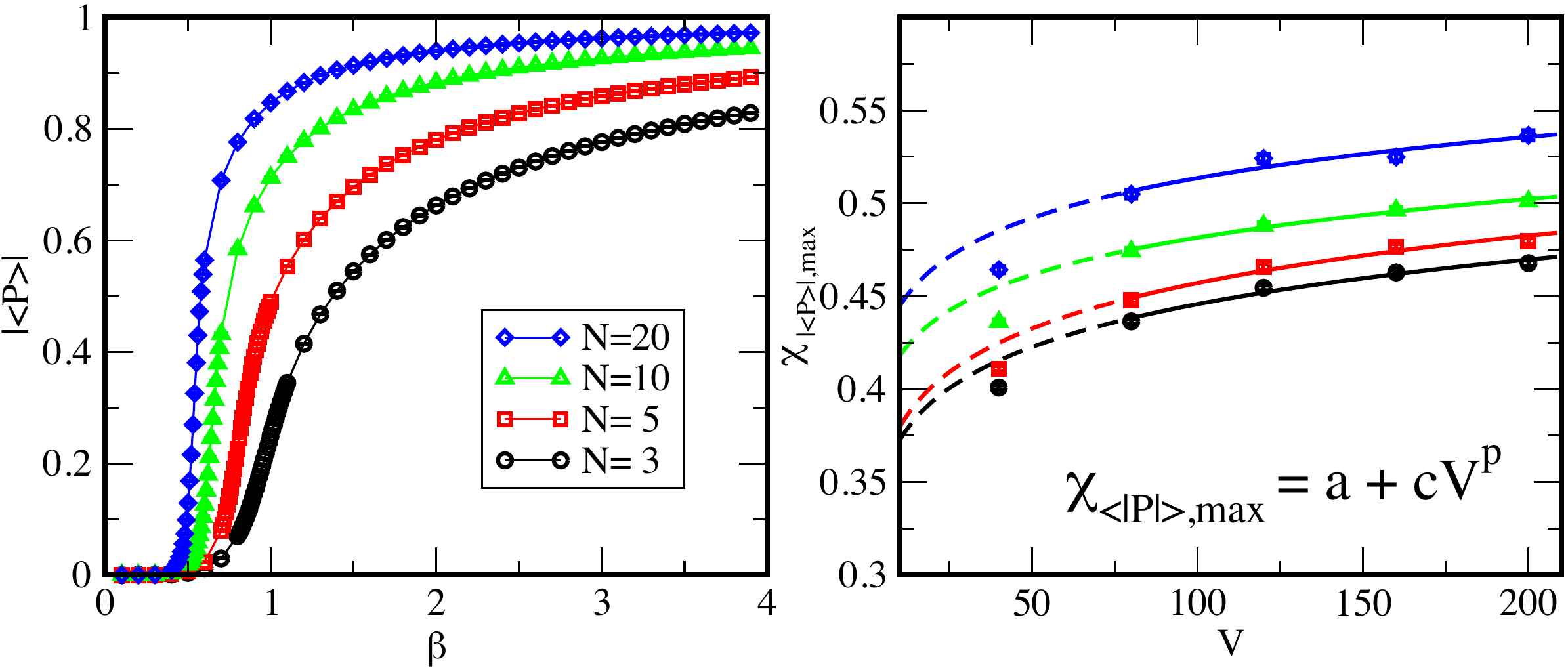}
  \includegraphics[width=0.3\linewidth]{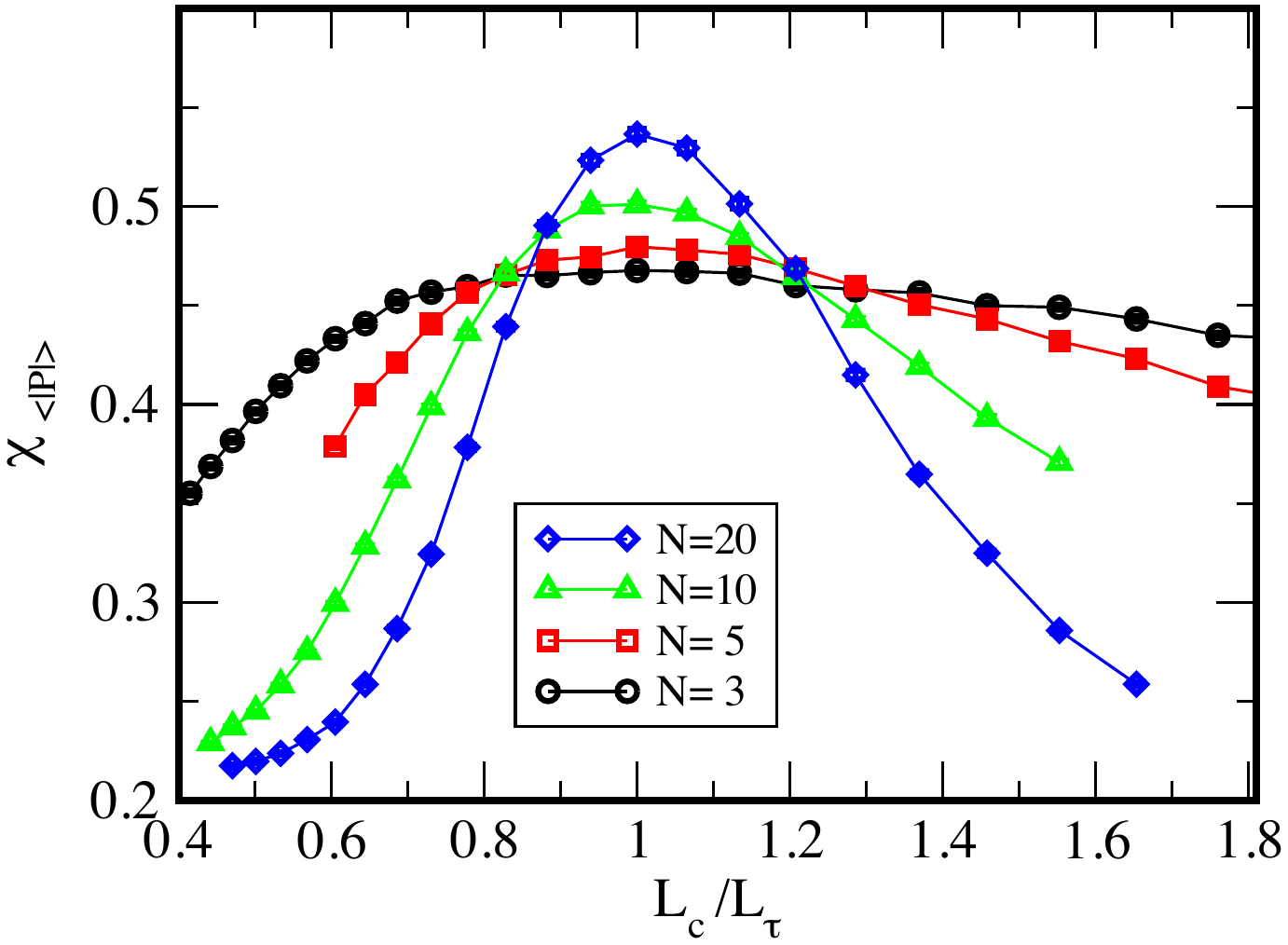}
 \caption{(Left) $|\langle P \rangle|$ as function of $\beta$ for PBC. 
(Center) Volume dependences of the maximal peak height of $\chi_{\langle | P| \rangle}$ for each $N$ by varying $N_{s}$ as $N_{s}=40,80,120,160,200$ with $N_{\tau}=8$ for PBC. 
(Right) $\chi_{\langle | P| \rangle}$ as a function of $L_{c}/L_{\tau}$ for PBC. These figures were shown in \cite{Fujimori:2019skd}.
 }
\label{fig:Poly}
\end{figure}

The Polyakov loop on the lattice is expressed as,
\begin{equation}
P\equiv {1\over{N_{s}}} \sum_{n_x}\prod_{n_\tau} \lambda_{n,\tau}.
\end{equation}
The distribution plots of Polyakov loop for $N=3$ at $\beta=0.1,1.0,2.0$ are depicted in Fig.~\ref{fig:dist},
where the Polyakov loops are distributed around the origin at low $\beta$ and gradually leave from the origin at higher $\beta$. 
The absolute values of the expectation values of Polyakov loop $|\langle P \rangle|$ are depicted as function of $\beta$ for $N=3,5,10,20$ and ($N_s,N_\tau$) $=$ ($200,8$) in the left panel of Fig.~\ref{fig:Poly}.
We find $|\langle P \rangle| \approx 0$ for low 
$\beta$ (large $L_{\tau}$) and $|\langle P \rangle| \not= 0$ for 
high $\beta$ (small $L_{\tau}$). For intermediate $\beta$, 
$|\langle P \rangle|$ gradually increases for small $N$.
These results strongly indicate a crossover behavior.

Next, we study the susceptibility of Polyakov loop. We here adopt that of the expectation values of absolute values of Polyakov loop $\chi_{\langle | P| \rangle}= V (\langle | P|^2 \rangle- \langle | P| \rangle^2)$ since the $\beta$ dependences of $|\langle P \rangle|$ and $\langle|P|\rangle$ are almost identical for PBC. 
In Fig.~\ref{fig:Poly} we show the volume dependence of the peak value of $\chi_{\langle | P| \rangle}$, by simulating the model with $N_{s}=40,80,120,160,200$ with $N_{\tau}=8$ fixed.  
We here fit the data points $N_s = 80,120,160,200$ by a function $\chi_{\langle|P|\rangle,{\rm max}} \,=\, a + c V^{p}\,$ with $V=N_s$.
The best fit values of the exponent are $p<1$ for $N=3,5,10,20$, which indicate the second-order or crossover transitions \cite{Fukugita:1990vu}. Furthermore, the ${\mathbb Z}_{N}$ symmetry is explicitly broken and the low and high temperature phases have the same symmetry. 
These facts imply that the transition is crossover.

We also show $\chi_{\langle | P| \rangle}$ as a function of $1/L_\tau$ in Fig.~\ref{fig:Poly} (Right), where the characteristic length $L_c$ for each $N$ is defined from the peak position of $\beta$ with ($N_s,N_\tau$)=($200,8$) fixed, then $\beta$ is translated into the length $L_\tau=N_\tau a$. The result indicates that the peak is broad for small $N$ but gets sharper as $N$ increases. This result again shows the crossover behavior for finite $N$ may be transformed into a conjectured phase transition in the large-$N$ limit \cite{Monin:2015xwa}.

\section{Polyakov loop for ${\mathbb Z}_{N}$-twisted boundary conditions}
\label{sec:TBC}
The ${\mathbb Z}_{N}$ intertwined symmetry between ${\mathbb Z}_{N}$ center and ${\mathbb Z_{N}}$ flavor-shift symmetries is exact in the ${\mathbb C}P^{N-1}$ model with ${\mathbb Z}_{N}$-TBC, where the ${\mathbb Z}_{N}$ symmetry is conjectured to be unbroken even at high temperature (small $L_{\tau}$) due to the transition between ${\mathbb Z}_{N}$ vacua via fractional instantons (see \cite{Dunne:2012ae,Misumi:2014jua,Fujimori:2016ljw} for the details).
We here perform the numerical Monte Carlo simulation for ${\mathbb Z}_{N}$-TBC on the small compactified direction $S_{\tau}^{1}$ for $N=3,5,10,20$ and $(N_{s}, N_{\tau})=(200,8)(400,12)$.
In this section, we show preliminary results of the distribution plot and the expectation value of Polyakov loop operator for this case, where one finds clear difference from the case with PBC.
We also exhibit configurations corresponding to fractional instantons and bions found in the simulation, 
which are also specific to ${\mathbb Z}_{N}$-TBC.

\subsection{Distribution plot}
\begin{figure}[t]
\centering
 \includegraphics[width=0.32\linewidth]{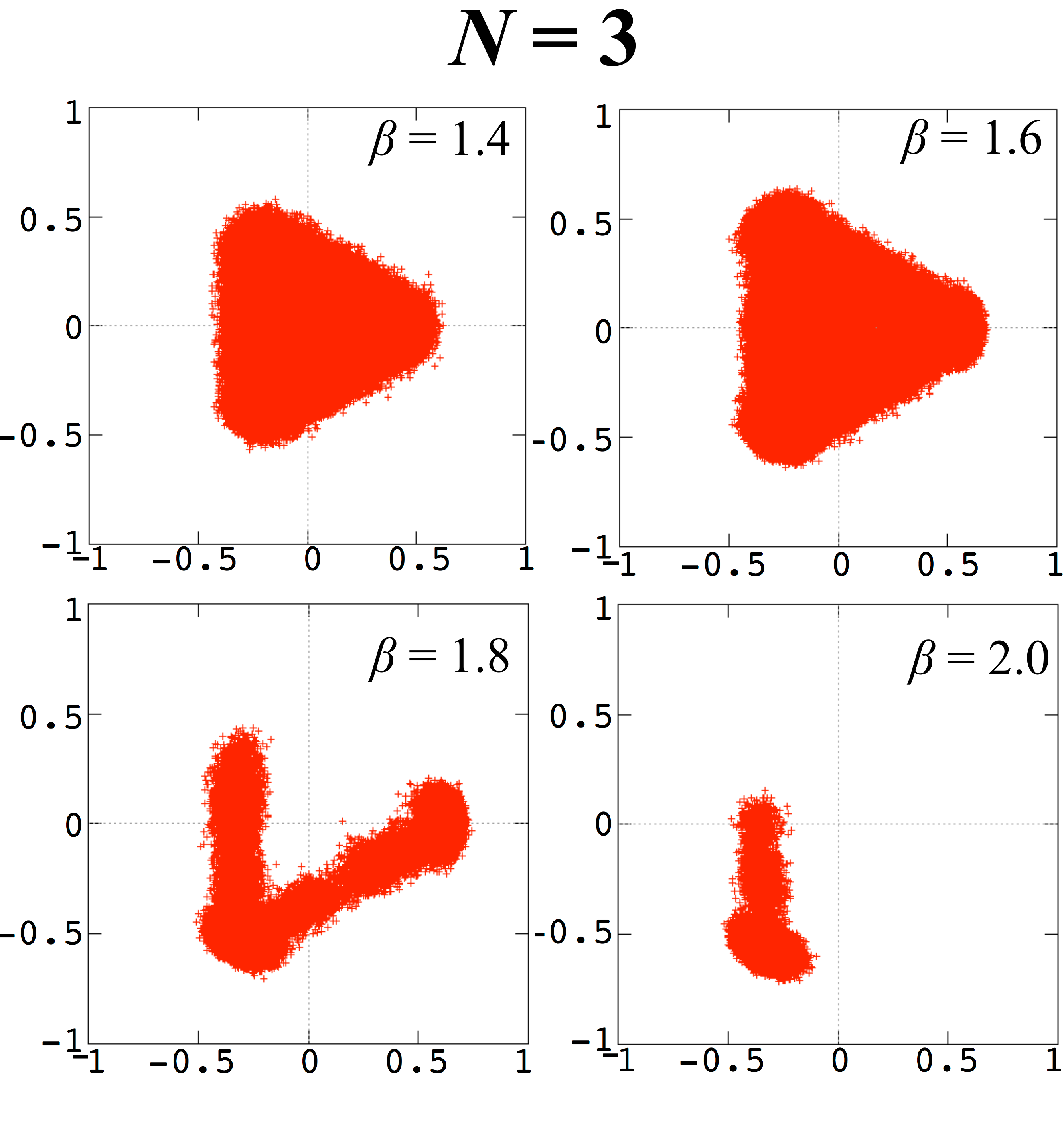}\,
  \includegraphics[width=0.32\linewidth]{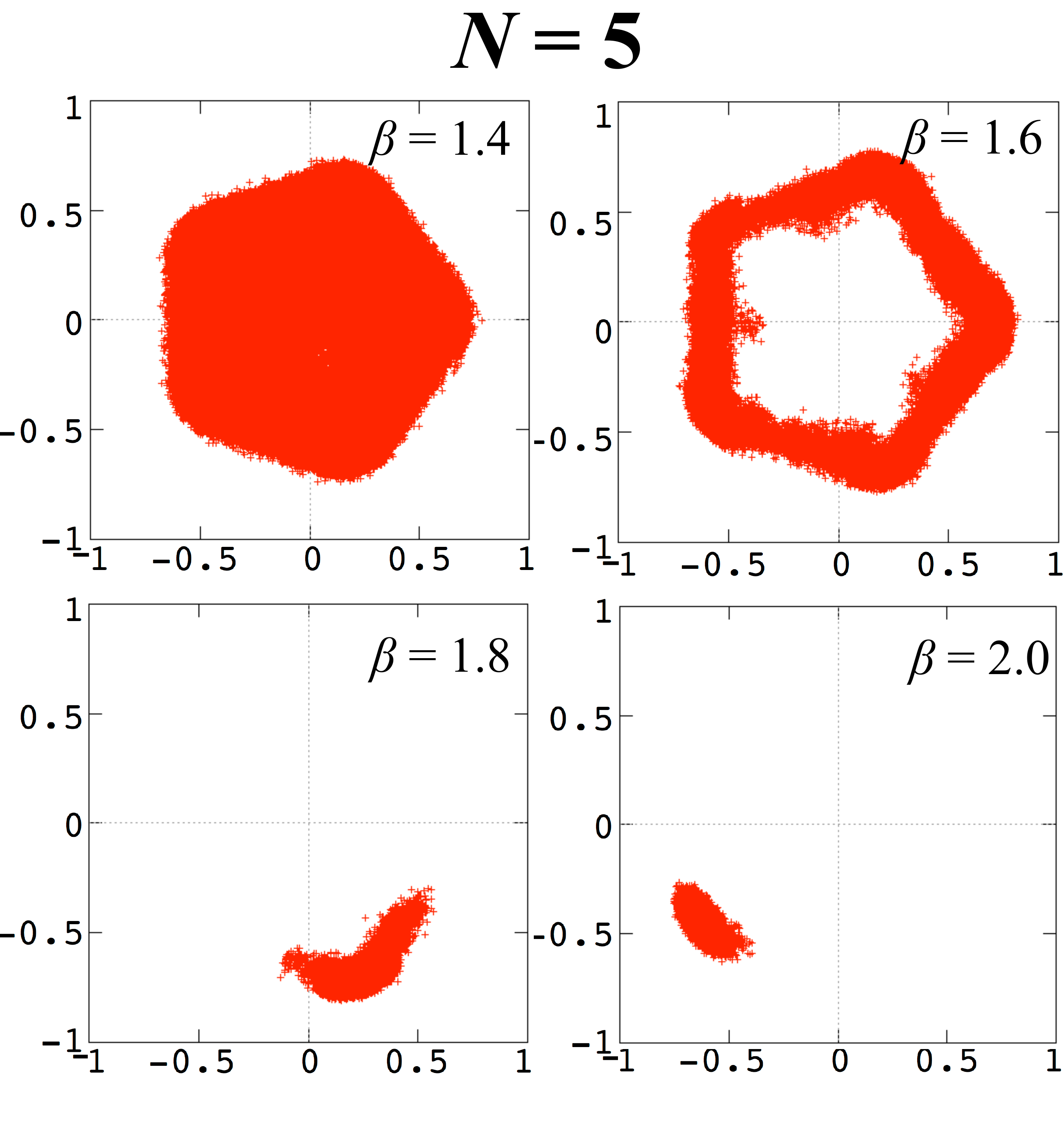}\,
   \includegraphics[width=0.32\linewidth]{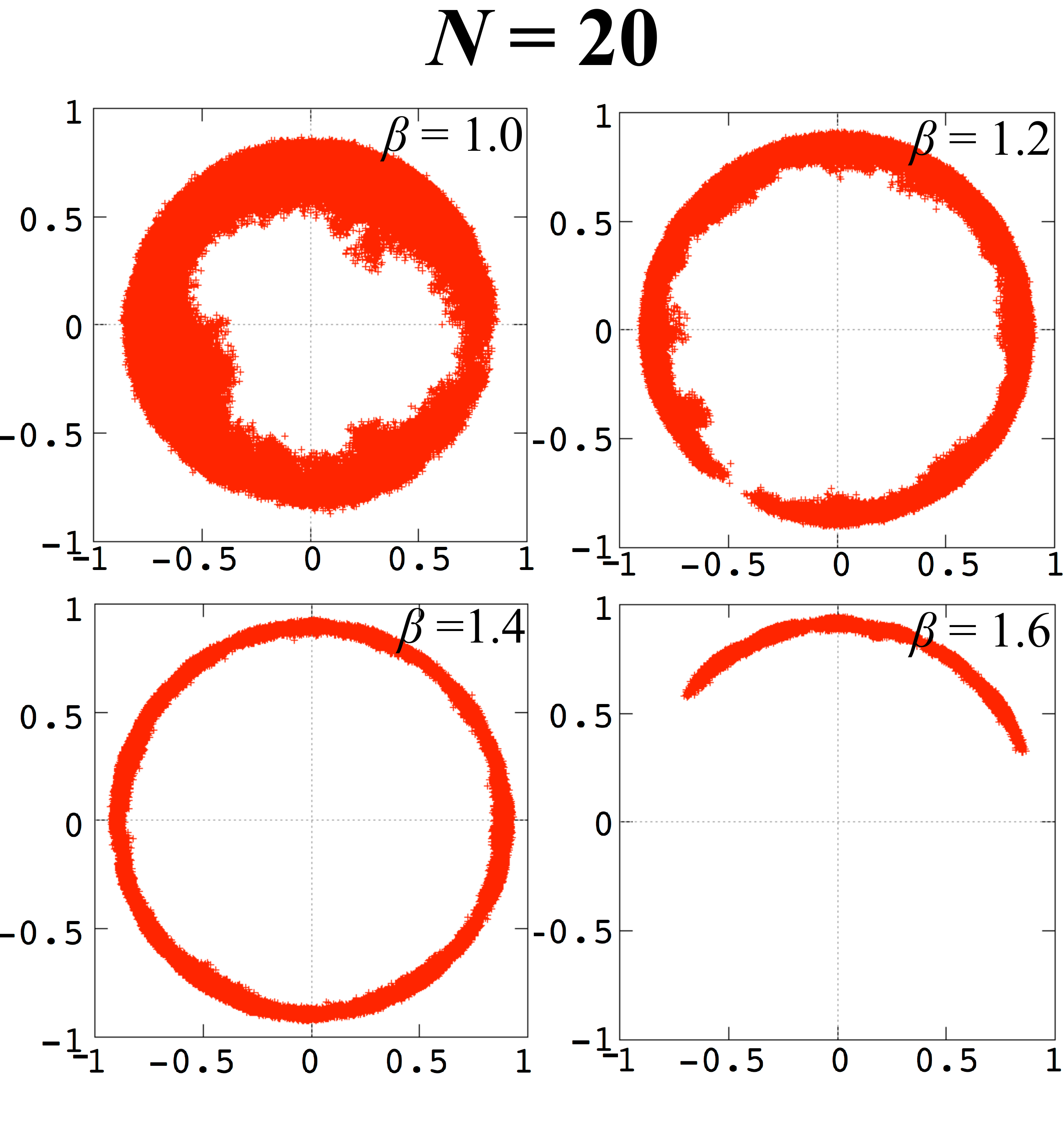}
 \caption{Distribution plots of Polyakov loop for $N=3,5,20$ with $(N_{s},N_{\tau})=(200,8)$ with ${\mathbb Z}_{N}$-TBC. Regular $N$-sided polygon shapes appear at intermediate $\beta$.}
\label{fig:dist_ZN}
\end{figure}

We note the Polyakov loops are distributed around the origin for low $\beta$ since the ${\mathbb Z}_{N}$ symmetry is exact for this case.
The question is how this distribution is altered for higher $\beta$ (small $L_{\tau}$).
In Fig.~\ref{fig:dist_ZN} we show the distribution plot of Polyakov loop with $\mathbb Z_{N}$-TBC from intermediate $\beta$ to high $\beta$ for $N=3,5,20$ with $(N_{s}, N_{\tau})=(200,8)$.
From low $\beta$ to intermediate $\beta$, the distributions gradually spread out and form regular $N$-sided polygons as seen in Fig.~\ref{fig:dist_ZN}.
This behavior implies unbroken ${\mathbb Z}_{N}$ symmetry with equivalent $N$ vacua, where one can speculate that there are nontrivial transitions among the $N$ vacua via fractional instantons.
For high $\beta$, the regular $N$-sided polygons are losing their shapes.
However, we speculate that this breaking of the polygon shapes could be an artifact originating in shortage of statistics
since $\beta$ at which the shape of polygons is broken gets larger by increasing the statistics $N_{\rm sweep}$.
We have checked this fact by comparing the results with different statistics $N_{\rm sweep} = 5000, 200000$.

\subsection{Expectation values of Polyakov loop}
\begin{figure}[t]
\centering
 \includegraphics[width=0.38\linewidth]{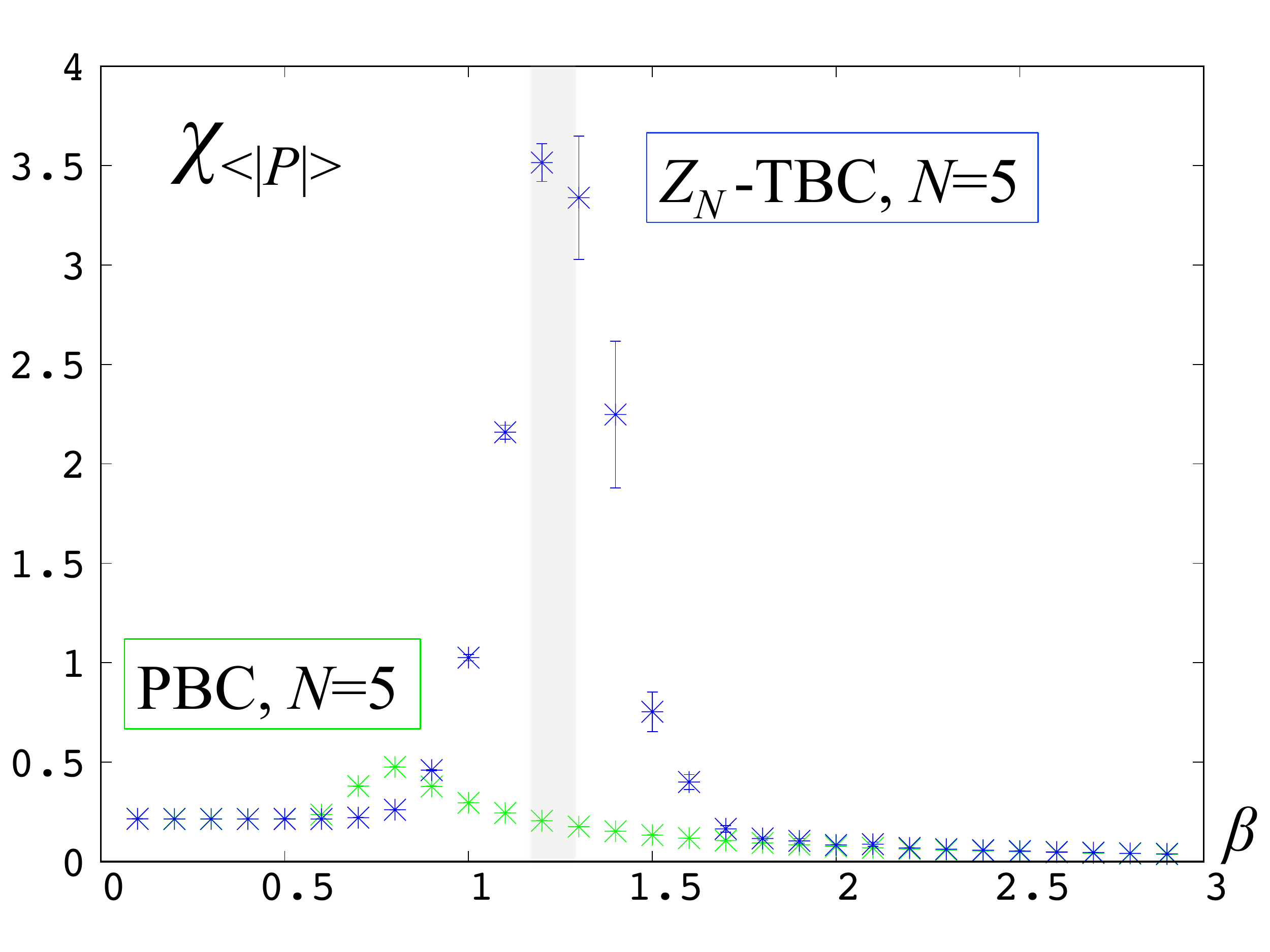}\,\,\,\,\,\,\,\,\,\,\,\,\,
 \includegraphics[width=0.38\linewidth]{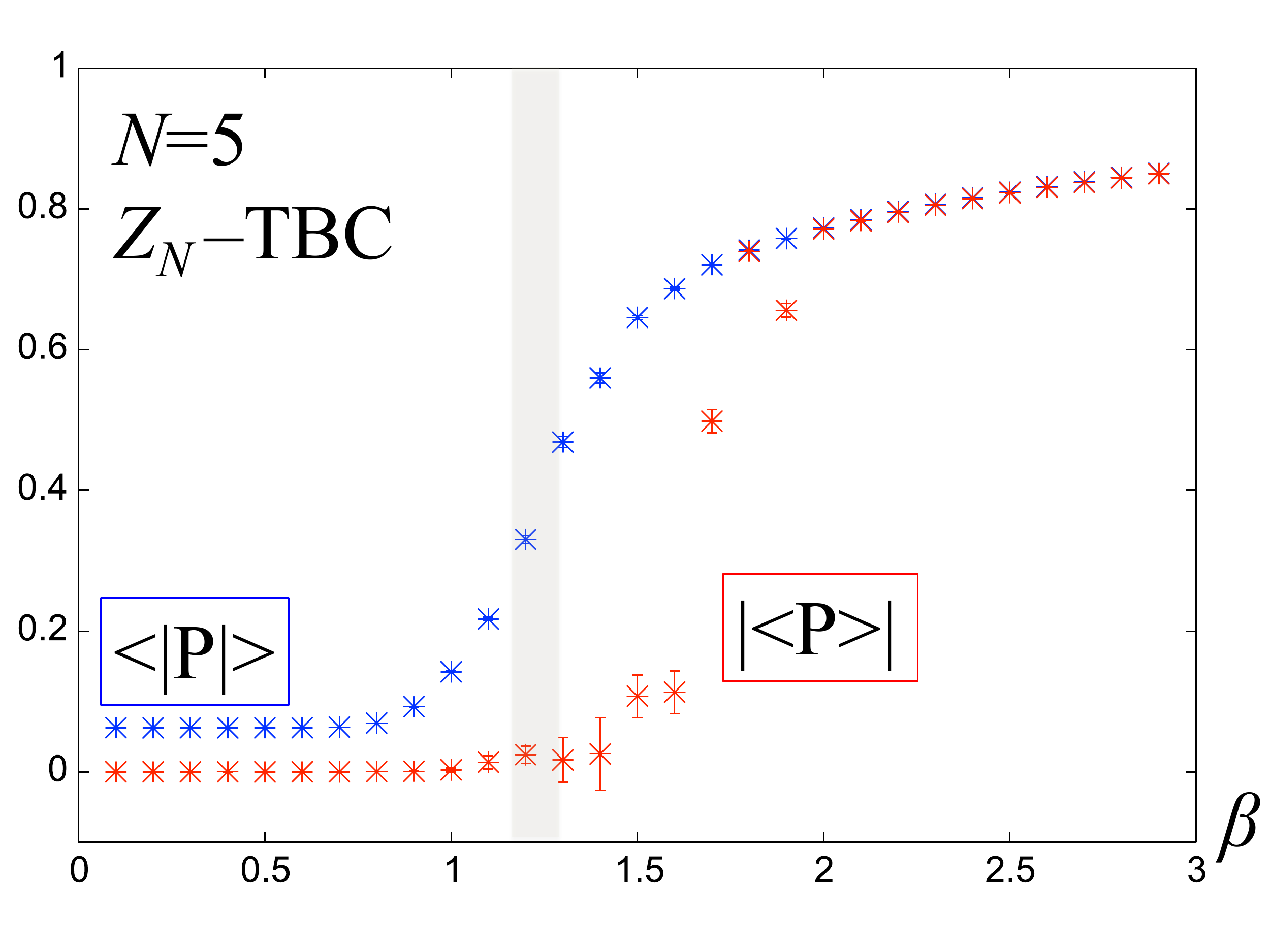}
 \caption{
(Left): Susceptibilities $\chi_{\langle |P| \rangle}$ of $\langle |P|\rangle$ for $\mathbb Z_{N}$-TBC and PBC with $N=5$.
(Right): Expectation values of Polyakov loop for $N=5$ with $(N_{s},N_{\tau})=(200,8)$ with ${\mathbb Z}_{N}$-TBC.
Red points indicate $|\langle P \rangle|$ while blue points $\langle |P|\rangle$. Grey lines with width indicate the characteristic $\beta$ read from the peak of $\chi_{\langle |P| \rangle}$.
}
\label{fig:Ploop_ZN}
\end{figure}

We show our results on expectation values of Polyakov loop by exhibiting results for $N=5$ as typical cases.
We note that similar results are obtained also for $N=3,10,20$.
Firstly, to compare scales between PBC and $\mathbb Z_{N}$-TBC cases, we depict susceptibilities of expectation values of absolute values of Polyakov loop $\chi_{\langle |P| \rangle}$ for $\mathbb Z_{N}$-TBC and PBC with $N=5$ in Fig.~\ref{fig:Ploop_ZN}(Left). 
The peak for ${\mathbb Z}_{N}$-TBC is located at higher $\beta$ (smaller $L_{\tau}$) than that for PBC, which indicates higher characteristic $\beta$ for ${\mathbb Z}_{N}$-TBC. The peak height itself is much higher than that of PBC.
We have to emphasize that {\it $\langle |P| \rangle$ is not an order parameter of ${\mathbb Z}_{N}$ symmetry}, but just indicates how broadly Polyakov loop distribution spreads. We speculate that this characteristic $\beta$ determined by the peak of $\chi_{\langle |P| \rangle}$ for ${\mathbb Z}_{N}$-TBC is the scale at which fractional instantons (bions) come to play an active role since fractional instanton configurations are found at high $\beta$ as we will show later.

Next, let us investigate a genuine order parameter of ${\mathbb Z}_{N}$ symmetry, that is the absolute values of expectation values of Polyakov loop $|\langle P \rangle|$.
In Fig.~\ref{fig:Ploop_ZN}(Right) we depict $|\langle P \rangle|$ for $\mathbb Z_{N}$-TBC, $N=5$ with $(N_{s}, N_{\tau})=(200,8)$ as red points.
For comparison to the characteristic scale, we also depict $\langle |P| \rangle$ by blue points.
For low $\beta$, the absolute values of expectation values of Polyakov loop $|\langle P \rangle|$ (red points) are consistent with zero, reflecting the exact ${\mathbb Z}_{N}$ symmetry.
For intermediate $\beta$, at which the distribution forms $N$-sided polygons, they are still small and highly fluctuate.
It is notable that the expectation values of absolute values of Polyakov loop $\langle |P|\rangle$ (blue points) become large at this intermediate $\beta$ since the distribution spread widely.
With a sufficient number of samples, we speculate that the values will get consistent with zero and the unbroken ${\mathbb Z}_{N}$ symmetry is realized by taking ensemble averages.
For high $\beta$, $|\langle P\rangle|$ suddenly increases and gets closer to unit.
These behaviors come from the breaking of the $N$-sided polygon shapes of the distribution at high $\beta$. 
However, as we have discussed above, we speculate that this phase-transition-like behavior is an artifact
since the critical $\beta$ for this transition gets larger by increasing the number of samples or enlarging the total volume.


\begin{figure}[t]
\centering
 \includegraphics[width=1.0\linewidth]{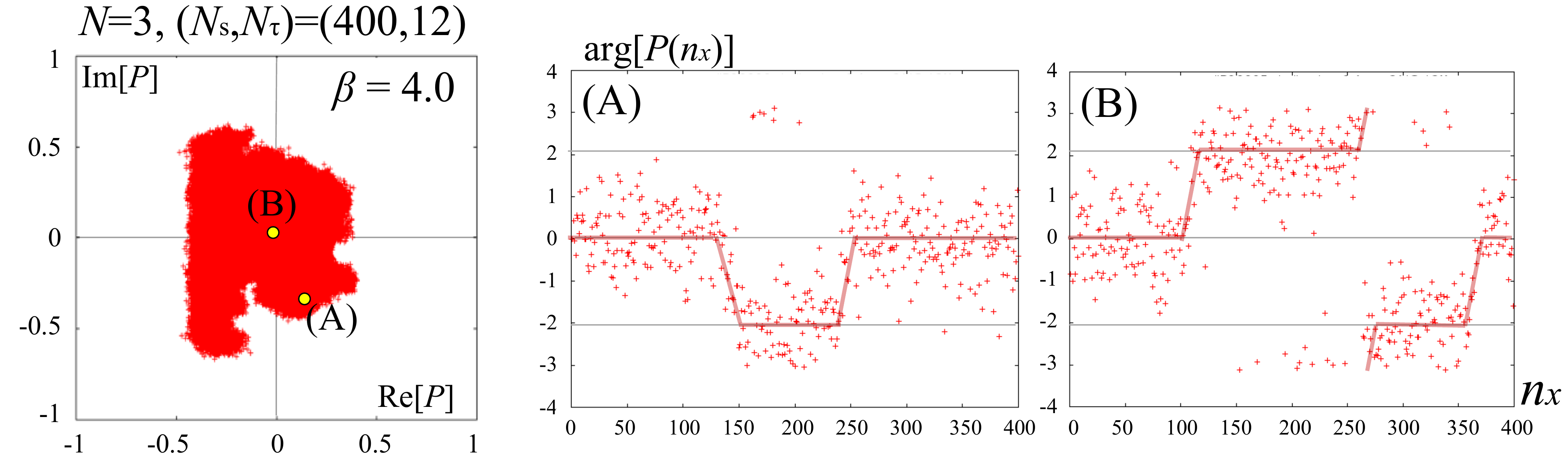}
 \caption{(Left): Distribution plot of Polyakov loop for $N=3$, $\beta=4.0$ with $(N_{s}, N_{\tau})=(400,12)$ for $\mathbb Z_{N}$-TBC.
(Center)(Right): Position dependences of ${\rm arg} [P(n_{x})]$ on $1\leq n_{x} \leq N_{s}$ for the two selected configurations (A)(B) in the distribution plot. We show the three vacua by gray lines and the speculated vacuum transitions by red lines.
(A) corresponds to a bion while (B) to three fractional instantons.}
\label{fig:frac_ZN}
\end{figure}

\subsection{Simulation on larger volume and fractional instantons}
We also obtain the simulation results with large volume $(N_{s}, N_{\tau})=(400,12)$ and
find a ${\mathbb Z}_{N}$ symmetric or regular $N$-sided polygon distribution even at quite high $\beta$.
We perform the simulation with $\beta=4.0$ and $N=3$, then obtain distribution plots of Polyakov loop for this parameter set. 
In Fig.~\ref{fig:frac_ZN}(Left) we depict one of the distributions plot of Polyakov loop.
It has a $N$-sided polygon shape (partly broken), which leads to a small expectation value of Polyakov loop.
We have checked that the history of ${\rm arg} [P]$ for this distribution is quite random, implying that the transition among ${\mathbb Z}_{N}$ occurs constantly. 
This result shows that, even at quite high $\beta$, the ${\mathbb Z}_{N}$-symmetric configuration exists with a certain probability ($\sim 5 \%$) for larger volume $(N_{s}, N_{\tau})=(400,12)$ while it is not the case for $(N_{s}, N_{\tau})=(200,8)$.

We next pick up two of configurations (A)(B) constituting the polygon-shaped distribution in Fig.~\ref{fig:frac_ZN}(Left) and investigate $n_{x}$ dependence of ${\rm arg} [P(n_{x})]$ with $P(n_{x})\equiv \prod_{n_\tau} \lambda_{n,\tau}$.
${\rm arg} [P(n_{x})] \approx \oint A_{\tau} d\tau$ describes the vacuum transition process since the topological charge $Q$ for ${\mathbb R}\times S^{1}$ is given by
\begin{equation}
Q= {1\over{2\pi}} \int \epsilon_{\mu\nu}\partial_{\mu}A_{\nu} = {1\over{2\pi}}
\left[
\oint A_{\tau}(x,\tau) d\tau \right]^{x=\infty}_{x=-\infty} .
\end{equation}
For our setup on $S_{s}^{1}$(large) $\times$ $S_{\tau}^{1}$(small), the periodic boundary condition is imposed on $n_{x}$ direction, thus the total $Q$ is zero but the transition process can be nontrivial.
In Fig.~\ref{fig:frac_ZN}(Center), ${\rm arg} [P(n_{x})]$ for a configuration located between the two adjacent
${\mathbb Z}_{3}$ vacua in Fig.~\ref{fig:frac_ZN}(Left) is depicted. 
In Fig.~\ref{fig:frac_ZN}(Right), ${\rm arg} [P(n_{x})]$ for a single configuration located near the origin in Fig.~\ref{fig:frac_ZN}(Left) is depicted.
The three classical vacua correspond to ${\rm arg} [P(n_{x})] =0, \pm 2\pi/3$ and we exhibit them by three gray lines in the figure.
The configuration in Fig.~\ref{fig:frac_ZN}(Center) is interpreted to be composed of one fractional instanton and one fractional anti-instanton (bion), while the configuration in Fig.~\ref{fig:frac_ZN}(Center) is interpreted to be composed of three fractional instantons constituting a single instanton.
Based on these results, we consider that the ${\mathbb Z}_{N}$-symmetric distribution of Polyakov loop is realized by the  ${\mathbb Z}_{N}$ vacuum transitions caused by the fractional instantons.
The bion configuration plays a pivotal role in the resurgent structure of the present model, whose contribution has nontrivial imaginary ambiguity canceled by that arising from the perturbative series.


\section{Summary and discussion}\label{sec:summary}
In the proceedings, we have reported the Monte Carlo simulations for 
the ${\mathbb C}P^{N-1}$ model on $S_{s}^{1}({\rm large}) \times S_{\tau}^{1}({\rm small})$:
for PBC in the $S_{\tau}^{1}$ direction, we have shown a confinement-deconfinement crossover by finding the continuous increase of the expectation value of Polyakov loop with $\beta$ being increased ($L_{\tau}=1/T$ being decreased).
For ${\mathbb Z}_{N}$-TBC,
we have found the $\beta$ dependence of expectation value of Polyakov loop completely differing from that for PBC.
Even for relatively high $\beta$, the $N$-sided polygon-shaped distributions of Polyakov loop produces the small expectation values, which imply the unbroken ${\mathbb Z}_{N}$ symmetry.
Although the polygons of the distribution lose their shapes for higher $\beta$,
we argue that it could be an artifact due to shortage of statistics or small volumes.
By adopting larger volumes, we have shown that the regular $N$-sided polygon-shaped distributions of Polyakov loop appear even for much higher $\beta$.
We have also argued the existence of the fractional instatons and bions by investigating the $n_x$-dependence of the Polyakov loop.

Our results give a novel understanding on the symmetry and the phase diagram of the $\C P^{N-1}$ model
on the compactified spacetime. 
In particular, the understanding of adiabatic continuity of ${\mathbb Z}_{N}$ symmetric phase is essential to study whether the resurgent structure in the $\C P^{N-1}$ model
on ${\mathbb R}\times S^{1}$ with ${\mathbb Z}_{N}$-TBC remains even in the decompactified limit.
We consider that our study in the present work will be the first step in this avenue.


\begin{thebibliography}{99}
  


\bibitem{Campostrini:1992ar} 
  M.~Campostrini, P.~Rossi and E.~Vicari,
  Phys.\ Rev.\ D {\bf 46}, 2647 (1992).
  
\bibitem{Farchioni:1993jd} 
F.~Farchioni and A.~Papa, 
Phys.\ Lett.\ B {\bf 306} 108 (1993). 
 
\bibitem{Fujimori:2019skd} 
  T.~Fujimori, E.~Itou, T.~Misumi, M.~Nitta and N.~Sakai,
  arXiv:1907.06925 [hep-th], to appear in PRD.
  
\bibitem{Monin:2015xwa} 
  S.~Monin, M.~Shifman and A.~Yung,
  Phys.\ Rev.\ D {\bf 92}, no. 2, 025011 (2015):{\bf 93}, no. 12, 125020 (2016).
    
\bibitem{Bolognesi:2019rwq} 
  S.~Bolognesi, S.~B.~Gudnason, K.~Konishi and K.~Ohashi,
  arXiv:1905.10555 [hep-th].

\bibitem{Dunne:2012ae} 
  G.~V.~Dunne and M.~\"{U}nsal,
  JHEP {\bf 1211}, 170 (2012).

\bibitem{Sulejmanpasic:2016llc} 
  T.\,Sulejmanpasic,
  Phys.\ Rev.\ Lett.\  {\bf 118}, no. 1, 011601 (2017).

\bibitem{Eto:2006mz} 
  M.\,Eto, et.al.,
 Phys.Rev.D {\bf 72} 025011 (2005);
  Phys.Rev.D {\bf 73}, 085008 (2006);
  J.Phys. A39 315 (2009).
  

\bibitem{Misumi:2014jua} 
  T.~Misumi, M.~Nitta and N.~Sakai,
  JHEP {\bf 1406}, 164 (2014);
  PTEP 2015 (2015) 033B02.

\bibitem{Fujimori:2016ljw} 
  T.\,Fujimori, et.al.,
  Phys.\ Rev.\ D {\bf 94}, no. 10, 105002 (2016);
JHEP {\bf 1902}, 190 (2019).

 \bibitem{Fukugita:1990vu} 
  M.~Fukugita, H.~Mino, M.~Okawa and A.~Ukawa,
  Phys.\ Rev.\ Lett.\  {\bf 65}, 816 (1990).



\end{thebibliography}
\end{document}